\documentclass[twocolumn,showpacs,preprintnumbers,prb,amsmath,amssymb]{revtex4}
\usepackage[dvips]{graphicx,color} 
\usepackage{dcolumn}% Align table columns on decimal point
\usepackage{bm}% bold math
\usepackage{subfigure} % Written by Steven Douglas Cochran
\begin{document}

%\preprint{APS/123-QED}

\title{Heterogeneous critical nucleation on a completely-wettable substrate}% Force line breaks with \\

\author{Masao Iwamatsu\footnote{iwamatsu@ph.ns.tcu.ac.jp}}
%\altaffiliation[Permanent address ]{Department of Physics, Tokyo City University, Setagaya-ku, Tokyo 158-8557, Japan}%Lines break automatically or can be forced with \\
%\email{iwamatsu@ph.ns.tcu.ac.jp}
\affiliation{
Department of Physics, Tokyo City University, Setagaya-ku, Tokyo 158-8557, Japan
}%
\date{\today}% It is always \today, today,

\begin{abstract}
Heterogeneous nucleation of a new bulk phase on a flat substrate can be associated with the surface phase transition called wetting transition.  When this bulk heterogeneous nucleation occurs on a completely-wettable flat substrate with a zero contact angle, the classical nucleation theory predicts that the free energy barrier of nucleation vanishes. In fact, there always exist a critical nucleus and a free energy barrier as the first-order pre-wetting transition will occur even when the contact angle is zero. Furthermore, the critical nucleus changes its character from the critical nucleus of surface phase transition below bulk coexistence (undersaturation) to the critical nucleus of bulk heterogeneous nucleation above the coexistence (oversaturation) when it crosses the coexistence.  Recently, Sear [J.Chem.Phys {\bf 129}, 164510 (2008)] has shown by a direct numerical calculation of nucleation rate that the nucleus does not notice this change when it crosses the coexistence.  In our work the morphology and the work of formation of critical nucleus on a completely-wettable substrate are re-examined across the coexistence using the interface-displacement model. Indeed, the morphology and the work of formation changes continuously at the coexistence. Our results support the prediction of Sear and will rekindle the interest on heterogeneous nucleation on a completely-wettable substrate. 

\end{abstract}

\maketitle

\section{\label{sec:sec1}Introduction}

The nucleation which occurs within the bulk metastable material is called {\it homogeneous nucleation}~\cite{Kelton10,Oxtoby88}.  However, it is frequently assisted by the presence of a surface.  The nucleation in this case is called {\it heterogeneous nucleation}~\cite{Kelton10,Oxtoby88,Turnbull49,Fletcher58,Gretz66,Scheludko81,Joanny86,Lazaridis93,Winter09}.  In the course of vapor to liquid nucleation, for example, a liquid droplet of semi-spherical shape will be formed on the substrate which attracts liquid within the {\it oversaturated} vapor.  The shape of this critical nucleus of the liquid is characterized by the apparent contact angle $\theta_{a}$ shown in Fig.~1.  

The nucleation is characterized by the nucleation rate $J$, which is the number of critical nuclei formed per unit time per unit volume.  Usually it is written in Arrhenius form
\begin{equation}
J = A \exp\left(-\frac{W}{k_{\rm B}T}\right)
\label{Eq:1x}
\end{equation}
where $A$ is a kinetic pre-exponential factor which is believed to be weakly dependent on temperature $T$, $k_{\rm B}$ is the Boltzmann's constant, and $W$ is the reversible work of formation of critical nucleus

According to the classical nucleation theory (CNT)~\cite{Turnbull49}, the work of formation (free energy barrier) of the heterogeneous nucleation $W_{\rm hetero}$ is expressed using the apparent contact angle $\theta_{a}$ and the free energy barrier $W_{\rm homo}$ of the homogeneous nucleation as 
\begin{equation}
W_{\rm hetero} = W_{\rm homo}f\left(\theta_{a}\right),
\label{Eq:2x}
\end{equation}
where
\begin{equation}
f\left(\theta_{a}\right)=\left(\theta_{a}-\cos\theta_{a}\sin\theta_{a}\right)/\pi
\label{Eq:3x}
\end{equation}
for the two-dimensional semi-cylindrical nucleus, and 
\begin{equation}
f\left(\theta_{a}\right)=\left(1-\cos\theta_{a}\right)^{2}\left(2+\cos\theta_{a}\right)/4.
\label{Eq:4x}
\end{equation}
for the three-dimensional axi-symmetric semi-spherical nucleus.  The contact angle $\theta_{a}$ characterizes the interaction between the liquid and the substrate.  When $\pi>\theta_{a}>0$ the liquid is said to incompletely wet the substrate.  When $\theta_{a}=0$, the liquid is said to completely wet the substrate. 
Since $0< f\left(\theta_{a}\right)<1$, the presence of the substrate enhances the nucleation rate (Eq.~(\ref{Eq:1x})) when the liquid incompletely wet the substrate ($\pi>\theta_{a}>0$) as $W_{\rm hetero}<W_{\rm homo}$.  When the liquid completely wet the substrate ($\theta_{a}=0$), CNT predicts that no nucleation barrier exist ($W_{\rm hetero}=0$) as $f\left(\theta_{a}=0\right)=0$.

\begin{figure}[htbp]
%Fig.1
\begin{center}
\includegraphics[width=1.0\linewidth]{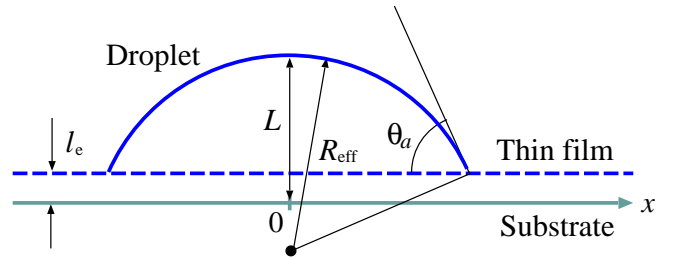}
\caption{
An ideal critical nucleus on a substrate.  The apparent contact angle $\theta_{a}$ is defined as the angle of the intersection of the cylindrical or spherical droplet surface with the effective radius $R_{\rm eff}$ and the height $L$ from the substrata and the wetting layer of the thickness $l_{e}$.
}
\label{fig:1}
\end{center}
\end{figure}

This heterogeneous nucleation of a new bulk phase on a flat substrate can be associated with the surface phase transition called wetting transition as both phase transitions can occur on the same surface~\cite{Talanquer96}.  In fact, there always exists a critical nucleus and a free energy barrier of the first-order so-called prewetting transition~\cite{Dietrich88,Bonn01,Bonn09} even on a completely-wettable substrate.  This transition occurs in the undersaturated vapor and separates states of thin and thick liquid films on a completely-wettable substrate with the zero contact angle. The shape of the critical nucleus is flat and pancakes-like~\cite{Joanny86b}, and cannot be characterized simply by a contact angle.  Therefore, in contrast to the prediction of CNT, the critical nucleus and the free energy barrier for the heterogeneous nucleation does exist even when the substrate is completely-wettable and the contact angle $\theta_{a}=0$ (Eq.~(\ref{Eq:2x})). By varying the chemical potential or the vapor pressure from undersaturation to oversaturation, the droplet changes its character from the critical nucleus of the prewetting surface phase transition to that of the vapor to liquid bulk phase transition.  Recently, Sear~\cite{Sear08} has shown by a direct numerical calculation of nucleation rate that the nucleus "does not notice" this change when it crosses the coexistence. 

In our work, the morphology and the work of formation of a critical nucleus rather than the nucleation rate~\cite{Sear08} on a completely-wettable substrate are re-examined across the coexistence using the interface-displacement model (IDM) which has been successfully used for studying wetting phenomena~\cite{Dietrich88,Bonn01,Bonn09}, line tensions~\cite{Indekeu92,Dobbs93} and even layering transitions~\cite{Iwamatsu98}.   In fact, various properties of critical nucleus of wetting layer rather than the bulk critical nucleus have already been studied.  For example, the stability and the critical exponents of the critical nucleus of wetting layer near the wetting and prewetting transitions have been studied using IDM~\cite{Bausch92,Bausch93,Blossey95,Bausch96} and Nakanishi-Fisher model~\cite{Nakanishi82,Blockhuis95}.  However those authors payed most attention to the nucleation of wetting transition in the undersaturated vapor {\it below} the bulk coexistence.  In fact, the heterogeneous nucleation of bulk phase transition occurs in the oversaturated vapor {\it above} the bulk coexistence.  Therefore, we will pay attention to those properties which is relevant to the bulk heterogeneous nucleation, which only a few number of authors such as Talanquer and Oxtoby~\cite{Talanquer96} and Sear~\cite{Sear08} have considered   

In Sec. II, we shall briefly recall the description the first-order surface phase transitions within the framework of IDM.  In Sec. III we shall study the morphology and the thermodynamics of critical droplets.  In particular, we shall focus our attention not to the scaling properties but on the relation between the heterogeneous nucleation and the prewetting transition.  Section IV is devoted to conclusion.

\section{A short review of the wetting transition and the interface-displacement model}

Within the interface displacement model (IDM) in $d$-dimensional apace the free energy of a fluid film of local thickness $l\left({\bf x}\right)$ is given by~\cite{Wyart91,Bauer99,Yeh99,Dobbs99,Starov09,Bausch92,Bausch93,Blossey95}
\begin{equation}
\Omega\left[l\right]=\int\left[\gamma \left(\left(1+ \left(\nabla l\right)^{2}\right)^{1/2}-1\right)+V\left( l\right)-\mu l\right]d^{d-1}x
\label{Eq:5x}
\end{equation}
where $\gamma$ is the liquid-vapor surface tension, $V\left(l\right)$ is the effective interface potential~\cite{Dietrich88,Bonn09,Dietrich91} from the substrate,  and $\mu$ denotes the deviation of the chemical potential from liquid-vapor coexistence such that $\mu=0$ is the bulk coexistence and for $\mu>0$ the vapor is oversaturated. In Eq.~(\ref{Eq:5x}) we keep the nonlinear dependence on $\nabla l$.  

\begin{figure}[htbp]
%Fig.1
\begin{center}
\includegraphics[width=1.0\linewidth]{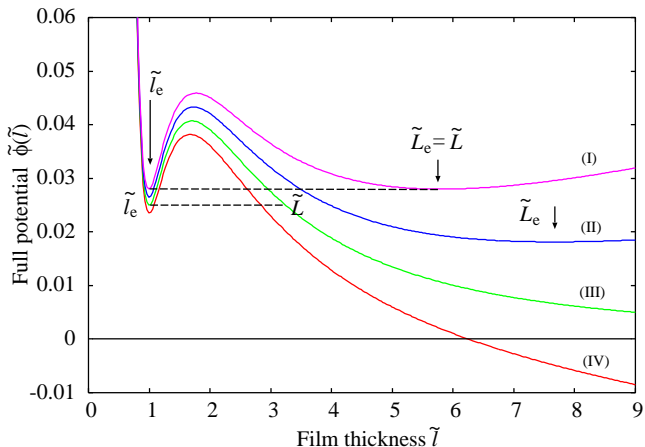}
\caption{
Reduced full potential $\tilde{\phi}(\tilde{l})$ (Eq.(\ref{Eq:23x})) as a function of the reduced film thickness $\tilde{l}$ (Eq.~(\ref{Eq:24x})) of a completely-wettable substrate for various reduced chemical potential $\tilde{\mu}$ (Eq.~(\ref{Eq:24x})). (I) at the prewetting line $\mu=\mu_{\rm p}$ ($\tilde{\mu}_{\rm p}=-0.00299$), (II) below the bulk coexistence and above the prewetting line $0>\mu>\mu_{\rm p}$ ($\tilde{\mu}=-0.0015$), (III) at the bulk coexistence $\mu=0$ ($\tilde{\mu}=0.0000$), (IV) above the bulk coexistence $\mu>0$ ($\tilde{\mu}=+0.0015$).  At the prewetting chemical potential, the thin (with thickness $l_{e}$) and thick (with thickness $L_{e}$) wetting film can coexist.  The thickness of the thick wetting film $L_{e}$ diverges above the bulk coexistence $\mu \ge 0$.  The $d=2$ dimensional droplet height $L$ is determined from the energy conservation law (\ref{Eq:18x}). 
}
\label{fig:2}
\end{center}
\end{figure}

Sometimes it is useful to consider the full potential $\phi(l)$ defined by
\begin{equation}
\phi(l)=V(l)-\mu l.
\label{Eq:6x}
\end{equation}
instead of the effective interface potential $V\left(l\right)$ and the chemical potential contribution $\mu l$ separately.  This full potential $\phi(l)$ depends not only on the chemical potential $\mu$ but also on the temperature $T$.  Figure \ref{fig:2} shows typical shapes of the full potential $\phi(l)$ of a completely-wettable substrate.  The full potential $\phi(l)$ exhibits double-well shape typical to the first-order surface phase transition, and its two minima at $l_{e}$ and at $L_{e}$ correspond to the metastable thin and the stable thick wetting films.  Figure~\ref{fig:2} indicates that the thick film become infinitely thick ($L_{e}=\infty$) when $\mu\ge 0$. 

\begin{figure}[htbp]
%Fig.1
\begin{center}
\includegraphics[width=0.9\linewidth]{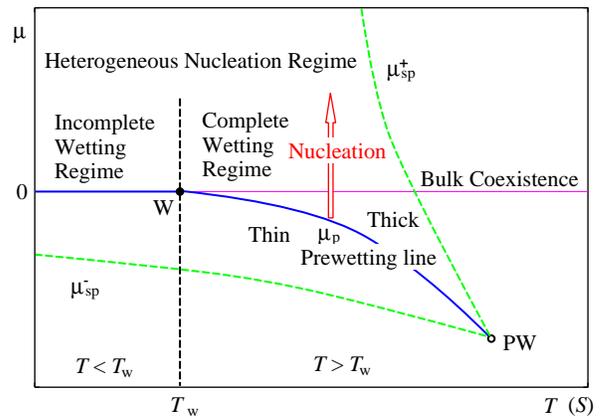}
\caption{
Schematic surface phase diagram in the spreading coefficient $S$ or the temperature $T$ and chemical potential $\mu$ plane.  The horizontal line $\mu=0$ corresponds to the bulk coexistence line, on which at the wetting point W with $S=0$ ($T=T_{\rm w}$) a first-order wetting transition occurs.  The prewetting transition line $\mu_{\rm p}$ starts at W and terminates at the prewetting critical point PW.  Above this prewetting line and below the bulk coexistence ($0>\mu>\mu_{\rm p}$), the stable state is a thick film.  The thickness diverges at the bulk coexistence and this infinitely thick film becomes bulk liquid phase above the coexistence ($\mu\ge 0$).  Below the prewetting line ($\mu<\mu_{\rm p}$), the stable state is a thin film.  A red arrow indicates the route in the complete-wetting regime where the isothermal heterogeneous nucleation is expected.}
\label{fig:3}
\end{center}
\end{figure}

The schematic surface phase diagram is shown in Fig.~\ref{fig:3} in the $T$-$\mu$ plane~\cite{Bonn01,Bonn09,Indekeu92,Dobbs93,Bausch92,Blossey95}.  Since the local minimum $V\left(l_{e}\right)$ at $l_{e}$ is related to the spreading coefficient $S$ and the temperature $T$ through~\cite{Bausch92,Blossey95}
\begin{equation}
\phi\left(l_{e}\right)=S\propto T-T_{\rm w},
\label{Eq:7x}
\end{equation}
where $T_{\rm w}$ is the wetting temperature, and $S$ is defined by
\begin{equation}
S=\gamma_{\rm sv}-\gamma_{\rm sl}-\gamma,
\label{Eq:8x}
\end{equation}
where $\gamma_{\rm sv}$ and $\gamma_{\rm sl}$ are the substrate-vapor and the substrate-liquid surface tensions, the first-order wetting transition from the incomplete wetting of a thin liquid film with thickness $l_{e}$ ($\phi\left(l_{e}\right)<0=\phi\left(l=\infty\right)$) to the complete wetting of a infinite thickness with $l=\infty$ ($\phi\left(l_{e}\right)>0=\phi\left(l=\infty\right)$) will occur at the wetting point W at $T=T_{\rm w}$ along $\mu=0^{-}$ in Fig.~\ref{fig:3}. Therefore, the complete-wetting regime with $S>0$ is realized above the wetting temperature $T>T_{\rm w}$.  Then, from the Young's formula
\begin{equation}
\gamma\cos\theta_{a}=\gamma_{\rm sv}-\gamma_{\rm sl}
\label{Eq:9x}
\end{equation}
we have
\begin{equation}
\cos\theta_{a}=1+\frac{S}{\gamma}.
\label{Eq:10x}
\end{equation}
Therefore the apparent contact angle vanishes ($\theta_{a}=0$) in the complete-wetting regime with $S\ge 0$ and, so does the free-energy barrier Eq.~(\ref{Eq:2x}).

Now, we will consider the surface phase diagram in the complete-wetting regime ($S>0, \;\;T>T_{\rm w}$).  In this regime, the prewetting transition $\mu_{\rm p}(T)$($<0$) line appears below the bulk coexistence line $\mu=0$ (Fig.~\ref{fig:3}). At the prewetting chemical potential $\mu=\mu_{\rm p}$, the full potential $\phi(l)$ has double-minimum shape with the same depth shown as the curve (I) in Fig.~\ref{fig:2}, which
\begin{equation}
\phi\left(l_{e}\right) = \phi\left(L_{e}\right),
\label{Eq:11x}
\end{equation}
and
\begin{equation}
\left.\frac{d\phi}{dl}\right|_{l_{e}} = \left.\frac{d\phi}{dl}\right|_{L_{e}}=0.
\label{Eq:12x}
\end{equation}
Then the thin (thickness $l_{e}$) and the thick (thickness $L_{e}$) wetting film can coexist at $\mu=\mu_{\rm p}<0$, which plays the role of the bulk coexistence $\mu=0$ for the heterogeneous nucleation in the complete-wetting regime ($T>T_{\rm w}$). When the chemical potential is increased ($0>\mu>\mu_{\rm p}$) along the arrow indicated in Fig.~\ref{fig:3}, the thick film with thickness $L_{e}$ becomes stable and the thin film with $l_{e}$ becomes metastable above the prewetting line  ($\phi\left(L_{e}\right)<\phi\left(l_{e}\right)$, curve (II) in Fig.~\ref{fig:3}). On the other hand, the thin film becomes stable and the thick film becomes metastable below the prewetting line ($\mu<\mu_{\rm p}$).  The metastable thin film loses stability at the upper spinodal $\mu=\mu_{\rm sp}^{+}$ and the metastable thick film loses stability at the lower-spinodal  $\mu=\mu_{\rm sp}^{-}$ shown in Fig.~\ref{fig:3}.  Finally, at and above the bulk coexistence $\mu\geq 0$ along the arrow in Fig.~\ref{fig:3}, the thickness of the stable thick film diverges (curves (III) and (IV) in Fig.~\ref{fig:3}).

Incidentally, previous authors payed most attention to the critical nucleus of wetting transition when $\mu<0$~\cite{Blockhuis95} and the critical phenomena near the critical point W and the prewetting line $\mu_{\rm p}$~\cite{Bausch92,Bausch93,Blossey95,Bausch96}.  We rather pay attention to the heterogeneous critical nucleus of bulk phase transition in a complete-wetting regime ($S>0, T>T_{\rm w}$) along the arrow indicated by "nucleation" in Fig.~\ref{fig:3} across the coexistence $\mu>0$ and $\mu\leq 0$.  In an incomplete-wetting regime ($S<0, T<T_{\rm w}$) with a finite contact angle $\theta_{a}>0$, a cylindrical and semi-spherical liquid droplet (Fig.~\ref{fig:1}) can appear on the substrate covered by a thin liquid film above the coexistence $\mu>0$.  This is in fact the critical nucleus of the heterogeneous nucleation and Eq.~(\ref{Eq:2x}) of CNT is qualitatively correct.

\section{Critical nucleus on a completely-wettable substrate}

\subsection{$d=2$-dimensional nucleus}
Based on the phase diagram shown in Fig.~\ref{fig:3} and the morphology of the potential $\phi(l)$ shown in Fig.~\ref{fig:2}, we can discuss the morphology and the work of formation of the critical nucleus on a completely-wettable substrate.  Since we are most interested in the global character of the heterogeneous nucleation in the complete-wetting regime, we will neglect the fluctuation effect that can be important near the wetting point W. 

For a $d=2$ dimensional cylindrical droplet, the Euler-Lagrange equation for the free energy (Eq.~(\ref{Eq:5x})) is simplified to~\cite{Bauer99,Yeh99,Dobbs99,Starov09}
\begin{equation}
\gamma \frac{d}{dx}\left(\frac{l_{x}}{\left(1+l_{x}^{2}\right)^{1/2}}\right)=\frac{dV}{dl}-\mu
\label{Eq:13x}
\end{equation}
or
\begin{equation}
\frac{\gamma l_{xx}}{\left(1+l_{x}^{2}\right)^{3/2}}=\frac{d\phi}{dl},
\label{Eq:14x}
\end{equation}
where $l_{xx}=d^{2}l/dx^{2}$.  Equation (\ref{Eq:14x}) could be considered as a kind of equation of motion for a classical particle moving in a potential $-\phi(l)$.

Equation (\ref{Eq:14x}) can be integrated once to give
\begin{equation}
\frac{-\gamma}{\left(1+l_{x}^{2}\right)^{1/2}}=-\gamma\cos\theta\left( l\right)=V\left(l\right)-\mu l+C,
\label{Eq:15x}
\end{equation} 
where $C$ is the integration constant and $\cos\theta\left( l\right)$ is the cosine of the angle $\theta(l)$ made between the tangential line of the liquid-vapor surface at the height $l(x)$ and the substrate~\cite{Yeh99}.  

Near the substrate, the liquid-vapor interface of the droplet will smoothly connect to the surrounding thin liquid film of thickness $l=l_{e}$ with $l_{x}=0$, the integration constant $C$ in Eq.~(\ref{Eq:15x}) will be given by $C=-\gamma-V\left( l_{e}\right)+\mu l_{e}$, and the liquid vapor interface will be determined from
\begin{equation}
\frac{-\gamma}{\left(1+l_{x}^{2}\right)^{1/2}}=\left(V\left(l\right)-V\left( l_{e}\right)\right)-\mu\left(l-l_{e}\right)-\gamma.
\label{Eq:16x}
\end{equation} 
Similarly, at the top of the droplet with a height $l=L$, we have again $l_{x}=0$ at $l=L$ (Fig.~\ref{fig:2}), and the liquid vapor interface will be determined from an equation similar to Eq.~(\ref{Eq:16x}) with $l_{e}$ replaced by $L$:
\begin{equation}
\frac{-\gamma}{\left(1+l_{x}^{2}\right)^{1/2}}=\left(V\left(l\right)-V\left( L\right)\right)-\mu\left(l-L\right)-\gamma.
\label{Eq:17x}
\end{equation}
Since Eqs.~(\ref{Eq:16x}) and (\ref{Eq:17x}) must be identical, we have
\begin{equation}
V\left(l_{e}\right)-\mu l_{e} = V\left( L\right)-\mu L,\;\;\;\mbox{or}\;\;\;\phi\left(l_{e}\right)=\phi\left(L\right),
\label{Eq:18x}
\end{equation}
which is similar to the energy conservation law for a classical particle whose (pseudo-)equation of motion is given by Eq.~(\ref{Eq:14x}) moving in a potential surface $-\phi(l)$.  Then, the height $L$ of the $d=2$ dimensional cylindrical droplet can be determined from Eq.~(\ref{Eq:18x}).  

Therefore a cylindrical droplet can exist on a completely-wettable substrate.  The height $L$ of the $d=2$ dimensional droplet is determined from Eq.~(\ref{Eq:18x}).  Also, it exists even in an undersaturated vapor below the bulk coexistence and above the prewetting line  ($0>\mu>\mu_{\rm p}$) in the thick film phase (Fig.~\ref{fig:3}). In this thick film phase, the droplet is not the critical nucleus of the bulk phase transition, but is in fact the {\it critical nucleus} of the thin-thick {\it surface phase transition}. This critical nucleus is expected to transform continuously into the bulk critical nucleus above the bulk coexistence ($\mu_{\rm sp}^{+}>\mu>0$) when the chemical potential cross the coexistence $\mu=0$.

In order to study the film thickness $l_{e}$, $L_{e}$ and the droplet height $L$ more quantitatively, we use a model interface potential
\begin{equation}
V\left(l\right)=V_{0}\left(\frac{1}{2}\left(\frac{l_{0}}{l}\right)^{2} - \frac{1+b}{3}\left(\frac{l_{0}}{l}\right)^{3} + \frac{b}{4}\left(\frac{l_{0}}{l}\right)^{4}\right)
\label{Eq:19x}
\end{equation}
where $l_{0}$ is a typical thickness of the thin-film, and $V_{0}>0$ plays the role of the so-called Hamaker constant~\cite{Israelachvili92,Iwamatsu98} $A_{\rm slv}$ of the substrate-liquid-vapor system through
\begin{equation}
\frac{V_{0}l_{0}^{2}}{2}=\frac{A_{\rm slv}}{12\pi},
\label{Eq:20x}
\end{equation}
and the parameter $b$ plays the role of the temperature that controls the transition from incomplete- to complete-wetting.  Since the two minima of Eq.~(\ref{Eq:19x}) locate at $l/l_{0}=1$ and $l/l_{0}=\infty$, and the spreading coefficient $S$ is related to the parameter $b$ through
\begin{equation}
S=V\left(l_{e}\right)=V_{0}\frac{2-b}{12},
\label{Eq:21x}
\end{equation}
the complete-wetting with $S>0$ is realized when $b<2$.  We will use the potential parameter $b=1.7$, which has already been used in Fig.~\ref{fig:2}.  

From Eq.~(\ref{Eq:19x}), the full potential (Eq.~(\ref{Eq:6x})) can be written as
\begin{equation}
\phi\left(l\right) = V_{0} \tilde{\phi}\left(\tilde{l}\right)
\label{Eq:22x}
\end{equation}
using non-dimensional reduced potential $\tilde{\phi}$ defined by
\begin{equation}
\tilde{\phi}\left(\tilde{l}\right)=\frac{1}{2\tilde{l}^{2}}-\frac{1+b}{3\tilde{l}^{3}}+\frac{b}{4\tilde{l}^{4}}-\tilde{\mu}\tilde{l},
\label{Eq:23x}
\end{equation}
and
\begin{equation}
\tilde{\mu}=\frac{\mu l_{0}}{V_{0}},\;\;\;\tilde{l} = \frac{l}{l_{0}}.
\label{Eq:24x}
\end{equation}
Figure~\ref{fig:2} shows the reduced potential $\tilde{\phi}\left(\tilde{l}\right)$ in the complete-wetting regime for various reduced chemical potentials $\tilde{\mu}$ below the bulk coexistence $\tilde{\mu}=0$ and above the prewetting $\tilde{\mu}_{\rm p}$.  We set $b=1.7$ for which the prewetting chemical potential is given by $\tilde{\mu}_{\rm p}=-0.00299326$.  

\begin{figure}[htbp]
\begin{center}
\includegraphics[width=1.0\linewidth]{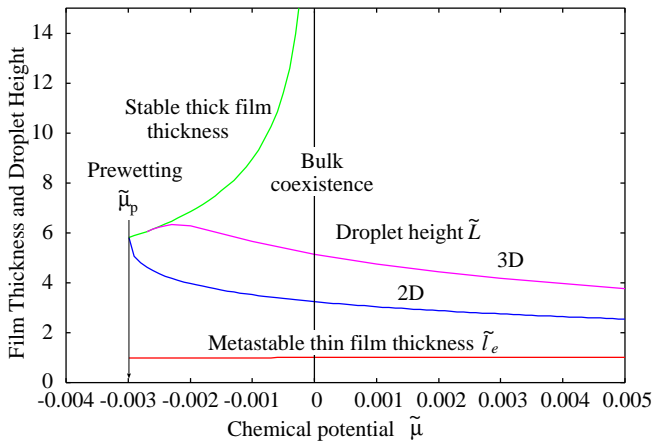}
\caption{
The stable thick-film thickness $\tilde{L}_{e}=L/l_{0}$ and the metastable thin-film thickness $\tilde{l}_{e}=l_{e}/l_{0}$ and the droplet height $\tilde{L}=L/l_{0}$ of the $d=2$ dimensional (2D) cylindrical nucleus determined from Eq.~(\ref{Eq:18x}).  The droplet height of the $d=3$ dimensional (3D) semi-spherical nucleus determined from the boundary value problem of the Euler-Lagrange equation (\ref{Eq:47x}) is also shown. }
\label{fig:4}
\end{center}
\end{figure}

Figure~\ref{fig:4} shows the reduced stable thick-film thickness $\tilde{L}_{e}=L_{e}/l_{0}$, the metastable thin-film thickness $\tilde{l}_{e}=l_{e}/l_{0}$, and the droplet height $\tilde{L}=L/l_{0}$ of the $d=2$ dimensional cylindrical nucleus determined from Eq.~(\ref{Eq:18x}). The height $L$ must be equal to the thick-film thickness $L_{e}$ and remains finite at the prewetting line ($\mu=\mu_{\rm p}$).  We observe that those quantities change continuously at the bulk coexistence ($\mu=0$) when the chemical potential $\mu$ is increased from negative ($\mu<0$, undersaturation) to positive ($\mu>0$, oversaturation). 

Using the expansion $V\left(l\right)-V\left( L\right)\simeq (l-L)\left.(dV/dl)\right|_{l=L}$, we have from Eq.~(\ref{Eq:17x})
\begin{equation}
\frac{-\gamma}{\left(1+l_{x}^{2}\right)^{1/2}}=\left(\left.\frac{dV}{dl}\right|_{l=L}-\mu\right)\left(l-L\right)-\gamma
\label{Eq:25x}
\end{equation}
near the top of the droplet $l\simeq L$.  If the effective chemical potential defined by
\begin{equation}
\mu_{\rm eff}=-\left.\frac{dV}{dl}\right|_{l=L}+\mu =-\left.\frac{d\phi}{dl}\right|_{l=L}
\label{Eq:26x}
\end{equation}
at the droplet top with height $L$ is positive, then Eq.~(\ref{Eq:17x}) becomes 
\begin{equation}
\frac{-\gamma}{\left(1+l_{x}^{2}\right)^{1/2}}=-\mu_{\rm eff}\left(l-L\right)-\gamma,
\label{Eq:27x}
\end{equation}
whose solution is a semi-circular shape (Fig.~\ref{fig:1})
\begin{equation}
l=\sqrt{R_{\rm eff}^2-x^2}-\left(R_{\rm eff}-L\right)
\label{Eq:28x}
\end{equation}
where $R_{\rm eff}-L$ is the shift of the base of circular interface.  The effective radius $R_{\rm eff}$ is given by the Kelvin-Laplace formula~\cite{Joanny86}
\begin{equation}
R_{\rm eff}=\frac{\gamma}{\mu_{\rm eff}}.
\label{Eq:29x}
\end{equation}

\begin{figure}[htbp]
%Fig.1
\begin{center}
\includegraphics[width=1.0\linewidth]{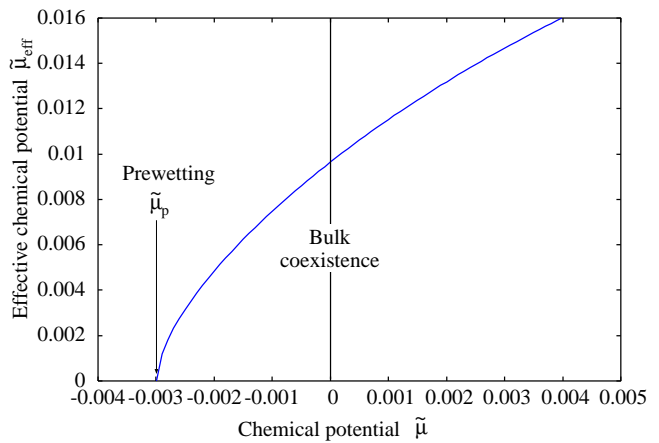}
\caption{
The effective chemical potentials $\tilde{\mu}_{\rm eff}=\mu_{\rm eff}l_{0}/V_{0}$ for the critical droplet in the complete-wetting regime ($b=1.7$). The effective chemical potential $\tilde{\mu}_{\rm eff}$ vanishes not at the bulk coexistence $\tilde{\mu}=0$, but at the prewetting line $\tilde{\mu}=\tilde{\mu}_{\rm p}$ below the bulk coexistence.  
}
\label{fig:5}
\end{center}
\end{figure}

In Fig.~\ref{fig:5}, we show the effective chemical potential $\tilde{\mu}_{\rm eff}=\mu_{\rm eff}l_{0}/V_{0}$  (Eq.~(\ref{Eq:26x})) in the complete-wetting regime as a function of the chemical potential $\tilde{\mu}$ calculated from the model potential (\ref{Eq:19x}) with $b=1.7$.  The effective chemical potential vanishes not at the bulk coexistence $\mu=0$, but at the prewetting line $\mu_{\rm p}$.  The effective circular liquid-vapor surface near the top of the nucleus can be maintained even in the undersaturated vapor with $\mu<0$ as far as $\mu_{\rm eff}>0$ (Eq.~(\ref{Eq:29x})).  The effective positive Laplace pressure ($\mu_{\rm eff}>0$) is produced by the repulsive surface potential ($\left.dV/dl\right|_{l=L}<0$) near the top of the droplet even though the vapor itself is undersaturated. 

At the prewetting $\mu=\mu_{\rm p}$, the effective chemical potential $\mu_{\rm eff}$ vanishes
\begin{equation}
\mu_{\rm eff}=-\left.\frac{dV}{dl}\right|_{l=L}+\mu_{\rm p}=0
\label{Eq:30x}
\end{equation}
from Eq.~(\ref{Eq:12x}) since $L=L_{e}$.  Hence, the prewetting line $\mu=\mu_{\rm p}$ acts as an effective or a shifted bulk coexistence with $\mu_{\rm eff}=0$ for the critical nucleus.  In fact, Blossey~\cite{Blossey95} has noted that the $d$-dimensional prewetting line corresponds to the $(d-1)$-dimensional bulk coexistence line. 

From Eqs.~(\ref{Eq:26x}) and (\ref{Eq:30x}), the effective chemical potential $\mu_{\rm eff}$ is approximately written as
\begin{equation}
\mu_{\rm eff}\simeq \left(1-\left(\left.\frac{dL}{d\mu}\right|_{\mu=\mu_{\rm p}}\right)\left(\left.\frac{d^{2}V}{dl^{2}}\right|_{l=L}\right)\right)\left(\mu-\mu_{\rm p}\right)
\label{Eq:31x}
\end{equation}
near the prewetting line $\mu=\mu_{\rm p}$, and the radius $R_{\rm eff}$ [Eq.~(\ref{Eq:29x})] of the curvature at the top of the droplets diverges as
\begin{equation}
R_{\rm eff}=\frac{\gamma}{\mu_{\rm eff}}\propto\frac{\gamma}{\mu-\mu_{\rm p}}
\label{Eq:32x}
\end{equation}
at the prewetting line $\mu=\mu_{\rm p}$.  This divergence has already been predicted by Bausch and Blossey~\cite{Bausch93,Blossey95} using a different definition of the droplet radius.  We note in Fig.~\ref{fig:5} that the effective chemical potential $\mu_{\rm eff}$ and, hence, the effective radius $R_{\rm eff}$ (Eq.~(\ref{Eq:29x})) at the top of the droplet change continuously at the bulk coexistence ($\mu=0$).

The lateral size and the liquid-vapor interface of the droplets can be studied only by solving the Euler-Lagrange equation (\ref{Eq:14x}), which can be done using the standard numerical method such as the Runge-Kutta method. To this end, we have to fix the parameter $V_{0}/\gamma$ which can be expressed by using the Hamakar constant $A_{\rm slv}$ in Eq.~(\ref{Eq:20x}) and $A_{\rm lvl}$ by
\begin{equation}
\frac{V_{0}}{\gamma}=-\frac{4D_{0}^{2}A_{\rm slv}}{l_{0}^{2}A_{\rm lvl}},
\label{Eq:33x}
\end{equation}
where an empirical formula~\cite{Israelachvili92}
\begin{equation}
\gamma = \frac{A_{\rm lvl}}{24\pi D_{0}^{2}}
\label{Eq:34x}
\end{equation}
with $D_{0}=0.165$nm is used. Suppose we tentatively set $l_{0}=2D_{0}$, and using the combining relation~\cite{Israelachvili92,Iwamatsu98}
\begin{eqnarray}
A_{\rm slv} &=& -\sqrt{A_{\rm ll}}\left(\sqrt{A_{\rm ss}}-\sqrt{A_{\rm ll}}\right),
\nonumber \\
A_{\rm lvl} &=& A_{\rm ll},
\label{Eq:35x}
\end{eqnarray}
we have
\begin{equation}
\frac{V_{0}}{\gamma} \sim \frac{A_{\rm slv}}{A_{\rm \rm lvl}} \sim \frac{\sqrt{A_{\rm ss}}-\sqrt{A_{\rm ll}}}{\sqrt{A_{\rm ll}}},
\label{Eq:36x}
\end{equation}
which will be $V_{0}/\gamma\sim 0.1-10$ using typical values of $A_{\rm ss}$ and $A_{\rm ll}$~\cite{Iwamatsu98}. By using the scaled quantities $\tilde{\mu}_{\rm eff}=\mu_{\rm eff}l_{0}/V_{0}$ and $\tilde{R}_{\rm eff}=R_{\rm eff}/l_{0}$, Eq.~(\ref{Eq:29x}) can be written as
\begin{equation}
\tilde{R}_{\rm eff} = \frac{1}{\left(V_{0}/\gamma\right)\tilde{\mu}_{\rm eff}}.
\label{Eq:37x}
\end{equation}
Therefore, the lateral size of nucleus which is roughly determined from $R_{\rm eff}$ is in inverse proportion to $V_{0}/\gamma$. 

\begin{figure}[htbp]
\begin{center}
\subfigure[The droplet shape when $V_{0}/\gamma=0.5$.]
{
\includegraphics[width=1.0\linewidth]{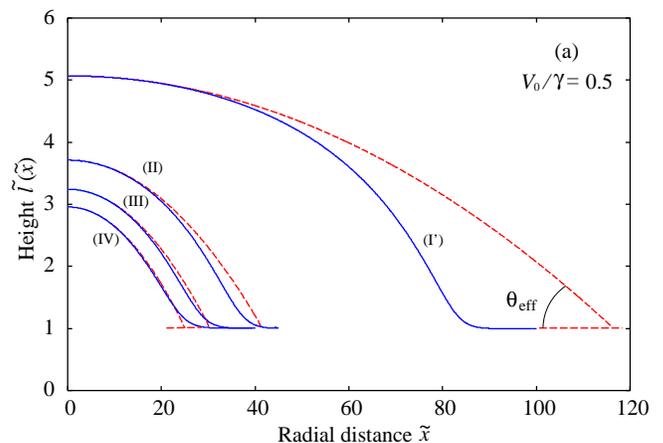}
%\vspace{3cm}
\label{fig:6a}}
\subfigure[The droplet shape when $V_{0}/\gamma=2.0$.]
{
\includegraphics[width=1.0\linewidth]{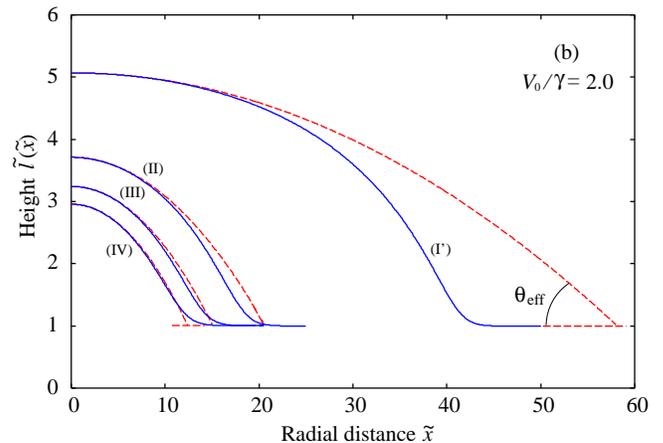}
%\vspace{3cm}
\label{fig:6b}}
\end{center}
\caption{
The droplet shape numerically determined from the Euler-Lagrange equation Eq.~(\ref{Eq:14x}) using the Runge-Kutta method (solid curves) compared with the ideal semi-circular shape given by Eq.~(\ref{Eq:28x}) (broken curves) (I') near the prewetting line ($\tilde{\mu}=-0.0029$), (II) below the bulk coexistence and above the prewetting line $0>\mu>\mu_{\rm p}$ ($\tilde{\mu}=-0.0015$), (III) at the bulk coexistence $\mu=0$ ($\tilde{\mu}=0.0000$), (IV) above the bulk coexistence $\mu>0$ ($\tilde{\mu}=+0.0015$).  Only a right half of the droplet is shown.  Note the scale of the vertical and the horizontal axes.  The droplet shape deviates significantly from circular shape and becomes "pancake" as the prewetting line is approached  $\mu\rightarrow\mu_{\rm p}$.} 
\label{fig:6}
\end{figure}

Figure \ref{fig:6} compares numerically determined $d=2$ dimensional cylindrical droplet shapes with the ideal semi-circular shapes (Eq.~(\ref{Eq:28x})) with the height $L$ and the effective radius $R_{\rm eff}$ calculated from Eqs.~(\ref{Eq:18x}) and (\ref{Eq:29x}) when $V_{0}/\gamma=0.5$ and $2.0$.  The droplet shape deviates significantly from an ideal circular shape, in particular, below the bulk coexistence $\mu<0$.  The droplet becomes flat and its shape becomes pancake-like~\cite{Joanny86b} as the chemical potential is decreased down to the prewetting line $\mu_{\rm p}$.  The size of the critical pancake is finite even at the prewetting line (Fig.~\ref{fig:6}) even though the effective radius $R_{\rm eff}$ diverges (Eq.~(\ref{Eq:32x})).  

It is possible to define an effective contact angle $\theta_{\rm eff}$ (see Fig.~\ref{fig:6}) of a droplet on a completely-wettable substrate by extrapolating the semi-circular shape (Eq.~(\ref{Eq:28x})) down to the surface of the thin film with thickness $l_{e}$. From the geometrical consideration shown in Fig.~\ref{fig:1}, we find
\begin{equation}
\cos\theta_{\rm eff}=1-\frac{L}{R_{\rm eff}}=1-\frac{\mu_{\rm eff}L}{\gamma}.
\label{Eq:38x}
\end{equation}
from Eq.~(\ref{Eq:29x}).  Therefore, the effective contact angle vanishes ($\theta_{\rm eff}\rightarrow 0$) at the prewetting line because $L\rightarrow L_{e}$ and $R_{\rm eff}\rightarrow \infty$ as $\mu\rightarrow \mu_{\rm p}$ from Eq.~(\ref{Eq:32x}).

The work of formation $W$ in Eq.~(\ref{Eq:1x}) can be calculated by inserting the droplet profile in Fig.~\ref{fig:6} obtained from the Euler-Lagrange equation (\ref{Eq:13x}) into Eq.~(\ref{Eq:5x}) with $d=2$ and subtracting the contribution from the wetting film with thickness $l_{e}$:
\begin{eqnarray}
W &=& \int\left[\gamma \left(\left(1+l_{x}^{2}\right)^{1/2}-1\right) \right. \nonumber \\
&+& \left. \left(V\left(l\right)-\mu l\right) - \left(V\left(l_{e}\right)-\mu l_{e}\right) \right]dx,
\label{Eq:39x}
\end{eqnarray}
which can be rewritten using Eq.~(\ref{Eq:16x}) as~\cite{Dobbs93}
\begin{eqnarray}
W &=& \int_{l=l_{e}}^{l=L} \frac{\gamma l_{x}}{\left(1+l_{x}^{2}\right)^{1/2}} dl,
\nonumber \\
&=& \int_{l=l_{e}}^{l=L} \left(2\gamma\Delta\phi\left(l\right)-\Delta\phi\left(l\right)^{2}\right)^{1/2} dl,
\label{Eq:40x}
\end{eqnarray}
where
\begin{equation}
\Delta\phi\left(l\right)=\phi\left(l\right)-\phi\left(l_{e}\right)=\phi\left(l\right)-\phi\left(L\right).
\label{Eq:41x}
\end{equation}
Equation (\ref{Eq:40x}) is also known as the {\it line tension}~\cite{Indekeu92,Dobbs93,Dobbs99}.  Here we interpret this energy as the work of formation of the critical nucleus on a completely-wettable substrate.  Intuitively, the critical nucleus in the complete-wetting regime is approximated by a thin flat disk (cf. Fig.~\ref{fig:6}) and its free energy is given only by the line tension of its perimeter.   

\begin{figure}[htbp]
\begin{center}
\subfigure[The work of formation in the complete-wetting regime.]
{
\includegraphics[width=1.0\linewidth]{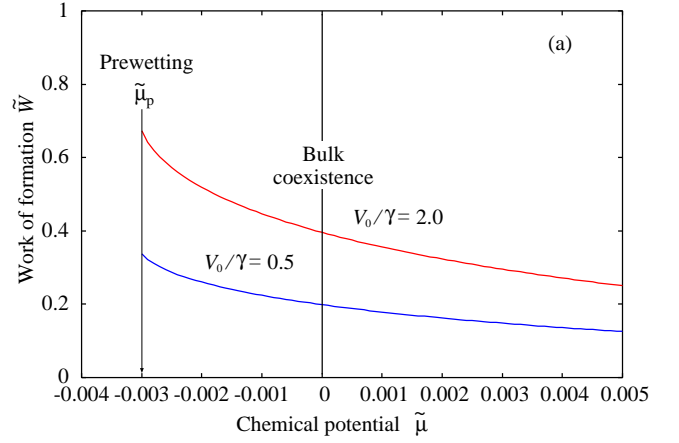}
%\vspace{3cm}
\label{fig:7a}}
\subfigure[The work of formation in the incomplete-wetting regime.]
{
\includegraphics[width=1.0\linewidth]{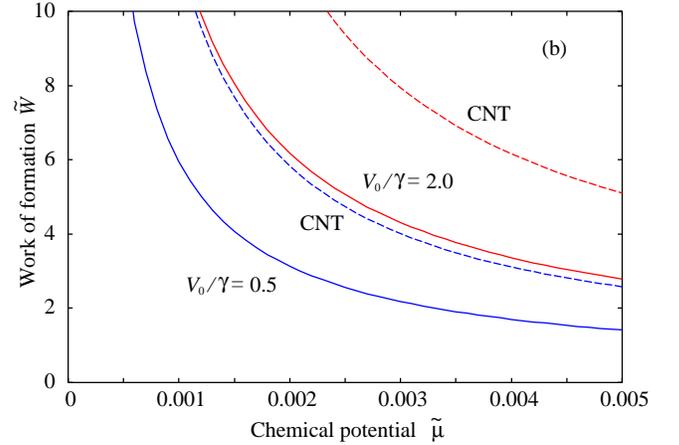}
%\vspace{3cm}
\label{fig:7b}}
\end{center}
\caption{
(a) The reduced work of formation $\tilde{W}=W/\gamma$ calculated from Eq.~(\ref{Eq:40x}) and the approximate formulas $\tilde{W}_{\rm CNT}=W_{\rm CNT}/\gamma$ calculated from Eqs.~(\ref{Eq:45x}) (a) in the complete-wetting and (b) in the incomplete-wetting regime. }
\label{fig:7}
\end{figure}

The reduced work of formation $\tilde{W}=W/\gamma$ from Eq.~(\ref{Eq:40x}) as a function of the reduced chemical potential $\tilde{\mu}$ is shown in Fig.~\ref{fig:7}(a). In this complete-wetting regime, the work of formation $W$ does not vanish and does not agree with the prediction of CNT.  Furthermore, it changes continuously even at the bulk coexistence at $\mu=0$ though the character of the droplet changes from the critical nucleus of the prewetting surface phase transition below bulk coexistence $\mu_{\rm p}<\mu<0$ to the critical nucleus of the heterogeneous bulk phase transition above the bulk coexistence $\mu>0$.  Therefore, the nucleation rate $J$ given by Eq.~(\ref{Eq:1x}) is expected to change continuously at the bulk coexistence as well~\cite{Sear08}.  A more detailed discussion on the continuity of the work of formation $W$ as well as that of the derivatives $d^{n}W/d\mu^{n}$ at $\mu=0$ will be given in the Appendix. 

This work of formation $W$ approaches a finite value at the prewetting line $\mu=\mu_{\rm p}$ (Fig.~\ref{fig:7}(a)) due to $d=2$ dimension as the integral (Eq.~(\ref{Eq:40x})) remains finite.  For such a flat nucleus (Fig.~\ref{fig:6}) with $l_{x}\ll 1$, we can approximate $\left(1+l_{x}^{2}\right)^{1/2}-1\simeq l_{x}^{2}/2$ in Eq.~(\ref{Eq:39x}), and Eq.~(\ref{Eq:40x}) can be approximated by~\cite{Indekeu92}
\begin{equation}
W \simeq \int_{l=l_{e}}^{l=L} \left(2\gamma\Delta\phi\left(l\right)\right)^{1/2} dl.
\label{Eq:42x}
\end{equation}
By using the scaled potential in Eq.~(\ref{Eq:23x}), we observe
\begin{equation}
W \propto \left(\gamma V_{0}\right)^{1/2}l_{0}.
\label{Eq:43x}
\end{equation}
Therefore, the reduced work of formation $\tilde{W}=W/\gamma$ is proportional to the parameter $\sqrt{V_{0}/\gamma}$ in Eq.~(\ref{Eq:33x})  (c.f. two curves in Fig.~\ref{fig:7}(a)).

Eq.~(\ref{Eq:40x}) can also be used to calculate the work of formation in the incomplete-wetting regime.  The result is shown in Fig.~\ref{fig:7}(b) when $b=2.5$ and $V_{0}/\gamma=0.5$ and $2.0$.   We also show the results using the CNT formula (\ref{Eq:2x}) and (\ref{Eq:3x}), which will be derived from Eq.~(\ref{Eq:40x}).  Using the definition of the local contact angle $\theta\left(l\right)$ in Eq.~(\ref{Eq:16x}), we can easily transform Eq.~(\ref{Eq:39x}) into simple and compact form
\begin{equation}
W = \gamma \int_{l=l_{e}}^{l=L} \sin\theta\left(l\right) dl.
\label{Eq:44x}
\end{equation}
By neglecting the surface potential $V(l)$ and using the change of variables $dl=\gamma\sin\theta d\theta/\mu$ from Eq.~(\ref{Eq:15x}), we have
\begin{eqnarray}
W_{\rm CNT} &=& \frac{\gamma^{2}}{\mu} \int_{\theta=\theta_{a}}^{\theta=\pi-\theta_{a}} \sin^{2}\theta d\theta,
\nonumber \\
&=& R_{\rm eff}^{2}\mu\left(\theta_{a}-\cos\theta_{a}\sin\theta_{a}\right),
\label{Eq:45x}
\end{eqnarray}
for the incomplete-wetting regime where we have used Eq.~(\ref{Eq:29x}) with $\mu_{\rm eff}=\mu$.  Equation (\ref{Eq:44x}) is exactly the work of formation in Eq.~(\ref{Eq:2x}) and (\ref{Eq:3x}) as $W_{\rm homo}=\pi R_{\rm eff}^{2}\mu$.  The apparent contact angle $\theta_{a}$ is calculated from the formula
\begin{equation}
\cos\theta_{a} = 1 + \frac{\mu\left(l_{e}-L\right)}{\gamma}
\label{Eq:46x}
\end{equation}
derived from Eqs.~(\ref{Eq:15x}) by neglecting the surface potential $V\left(l\right)$.

The work of formation $W$ calculated from Eq.~(\ref{Eq:40x}) is always smaller than the work of formation $W_{\rm CNT}$ calculated from the classical formula (\ref{Eq:45x}), which is due to the presence of the attractive surface potential $V(l)$ which enhances the condensation of vapor and lower the barrier of nucleation.  Both $W$ and $W_{\rm CNT}$ diverges at the bulk coexistence $\mu=0$.  Therefore, CNT is qualitatively correct in the incomplete-wetting regime.  

The work of formation in the complete-wetting regime in Fig.~\ref{fig:7}(a) is smaller than that in the incomplete-wetting regime in Fig.~\ref{fig:7}(b).  However, the former quantity is still appreciable and as far as $W/k_{\rm B}T>76$ the free energy barrier of nucleation should be observable~\cite{Oxtoby88} even in the complete-wetting regime.  Also, this free energy barrier and, therefore, the nucleation rate calculated from Eq.~(\ref{Eq:1x}) should change continuously~\cite{Sear08} at the bulk coexistence.

\subsection{$d=3$ dimensional nucleus}
The profile of the droplet on the substrate is determined by minimizing the free energy (Eq.~(\ref{Eq:5x})) with respect to the film thickness $l$ which leads to the Euler-Lagrange equation.  For a $d=3$ dimensional hemispherical droplet, it is given by~\cite{Dobbs99}
\begin{equation}
\gamma \left(\frac{d}{dx}+\frac{1}{x}\right)\left(\frac{l_{x}}{\left(1+l_{x}^{2}\right)^{1/2}}\right)=\frac{dV}{dl}-\mu,
\label{Eq:47x}
\end{equation}
where $l_{x}=dl/dx$ and $x$ is the coordinate measured from the center of the base of the droplet (Fig.~\ref{fig:1}). This equation cannot be integrated to give Eq.~(\ref{Eq:13x}), as for the $d=2$ dimensional cylindrical droplet. As a result, it is not possible to calculate the height $L$ of the nucleus from Eq.~(\ref{Eq:16x}).  Instead, the height $L$ is determined from the solution of Eq.~(\ref{Eq:47x}) that satisfies the boundary condition $l_{x}=0$ at $x=0$ and $l=l_{e}$ at $x=\infty$.

\begin{figure}[htbp]
\begin{center}
\includegraphics[width=1.0\linewidth]{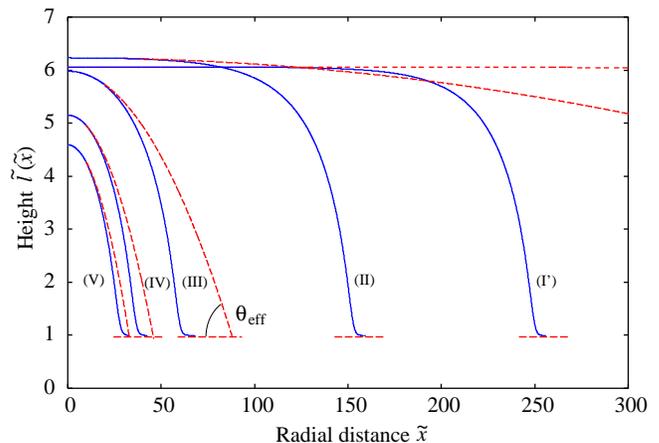}
%\vspace{3cm}
\end{center}
\caption{
The droplet shape numerically determined from the Euler-Lagrange equation (\ref{Eq:47x}) using the Runge-Kutta method (solid curves) when $V_{0}/\gamma=2.0$ compared with the ideal semi-spherical shape given by Eq.~(\ref{Eq:28x}) (broken curves) for (I') near the prewetting line ($\tilde{\mu}=0.0027$), (II) below the bulk coexistence and above the prewetting line $0>\mu>\mu_{\rm p}$ ($\tilde{\mu}=-0.0025$), (III) $\tilde{\mu}=-0.0015$, (IV) at the bulk coexistence $\mu=0$ ($\tilde{\mu}=0.0000$), (V) above the bulk coexistence $\mu>0$ ($\tilde{\mu}=+0.0015$).  The droplet shape cannot be determined near the spinodal as its lateral size diverges.  Only a right half of the droplet is shown.  Note the scale of the vertical and the horizontal axes.} 
\label{fig:8}
\end{figure}

In Fig.~\ref{fig:8} we show numerically determined droplet shapes for various $\mu$ when $V_{0}/\gamma=2.0$ .  In Fig.~\ref{fig:4} we have shown the droplet height $L$ directly determined from the numerically determined droplet shape in Fig.~\ref{fig:8} as the function of the chemical potential $\mu$.  By repeating the argument from Eq.~(\ref{Eq:25x}) to (\ref{Eq:29x}), it is possible to approximate the droplet shape determined from Eq.~(\ref{Eq:47x}) near the top of the droplet by the semi-spherical shape given by Eq.~(\ref{Eq:28x}).  However, the effective radius is now given by
\begin{equation}
R_{\rm eff} = \frac{2\gamma}{\mu_{\rm eff}}
\label{Eq:48x}
\end{equation}
instead of Eq.~(\ref{Eq:29x}).  Then, it is possible to define the effective contact angle $\theta_{\rm eff}$ for a $d=3$ dimensional droplets on a completely-wettable substrate shown in Fig.~\ref{fig:8}. This effective contact angle is also given by Eq.~(\ref{Eq:38x}) and vanishes ($\theta_{\rm eff}\rightarrow 0$) at the prewetting line because $R_{\rm eff}\rightarrow \infty$ as $\mu\rightarrow \mu_{\rm p}$.

In contrast to the $d=2$ dimensional case, the lateral size of the $d=3$ dimensional droplet is larger and diverges at the prewetting line $\mu=\mu_{\rm p}$. Using an analogy to the classical mechanics, this divergence can be intuitively understandable.  Since the Euler-Lagrange equation (\ref{Eq:47x}) can be interpreted as an equation of motion of a classical particle~\cite{Bausch92} in a potential field $-\phi(l)$ and the term $(l_{x}/x)$ plays the role of friction, it takes infinitely long "time" $x$ for a classical particle starting from the one maximum of the potential $-\phi(l)$ at $L=L_{e}$ to reach the another maximum with the same height at $l_{e}$ (see Fig.~\ref{fig:3} at the prewetting when $\tilde{\mu}_{\rm p}=-0.00299$).  Therefore the "lateral size" $x$ of the $d=3$ dimensional critical droplet diverges at the prewetting line.  For a $d=2$ dimensional droplet, since there is no friction term and the energy conservation is assured from Eq.~(\ref{Eq:18x}), the lateral size of the droplet remains finite at the prewetting.   

\begin{figure}[htbp]
\begin{center}
\includegraphics[width=1.0\linewidth]{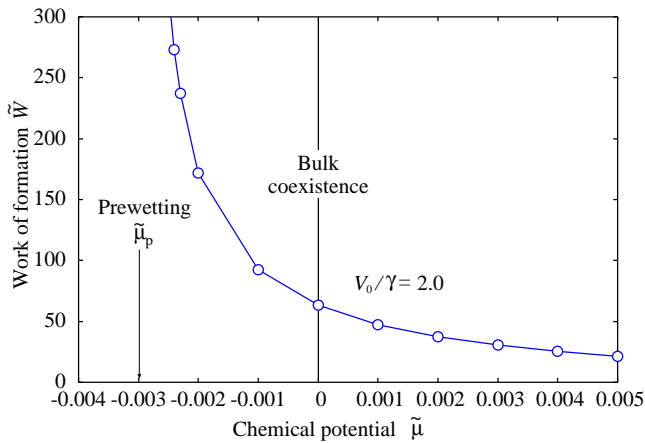}
%\vspace{3cm}
\end{center}
\caption{
The reduced work of formation $\tilde{W}=W/\gamma$ calculated from Eq.~(\ref{Eq:5x}) by using the droplet profile $l(x)$ numerically determined from Eq.~(\ref{Eq:47x}) when $V_{0}/\gamma=2.0$ as a function of the reduced chemical potential $\tilde{\mu}$.  In contrast to the $d=2$ dimensional nucleus, the work of formation $W$ of the $d=3$ dimensional nucleus diverges at the prewetting $\mu_{\rm p}$. }
\label{fig:9}
\end{figure}

Since we can not integrate Eq.~(\ref{Eq:47x}) analytically for a $d=3$ dimensional droplet, we cannot use Eq.~(\ref{Eq:40x}) to calculate the work of formation $W$.  Instead, we have to resort to a direct numerical integration using Eq.~(\ref{Eq:5x}).  Figure \ref{fig:9} shows the reduced work of formation $\tilde{W}=W/\gamma$ of a $d=3$ dimensional critical nucleus as a function of the reduced chemical potential $\tilde{\mu}$.

Similar to the $d=2$ dimensional critical nucleus, the work of formation $W$ is a continuous function of the chemical potential $\mu$ at the bulk coexistence. However, in contrast to the $d=2$ dimensional nucleus in Fig.~\ref{fig:7}(a), the work of formation $W$ of the $d=3$ dimensional nucleus diverges at the prewetting $\mu_{\rm p}$ in Fig.~\ref{fig:9}.

\section{\label{sec:sec4}Conclusion}
In this paper, we have used the interface displacement model to study the heterogeneous nucleation on a completely-wettable planar substrate where the first order incomplete to complete wetting surface phase transition takes place.  The prediction of the classical nucleation theory (CNT) is critically tested.  It is found that the classical picture breaks down on a completely-wettable substrate, where CNT predicts that the critical nucleus and the nucleation barrier are expected to vanish since the apparent contact angle vanishes.  In fact, both the critical nucleus of the heterogeneous bulk phase transition in the oversaturated vapor and the critical nucleus of the surface thin-thick prewetting transition in the undersaturation vapor can exist and transform continuously at the bulk coexistence on a completely-wettable substrate. 

Therefore, the critical nucleus as well as the free-energy barrier of nucleation exist on a completely-wettable substrate.  Furthermore, the nucleus exists even under the undersaturated vapor.  The undersaturated vapor turns to an effectively oversaturated vapor due to the effective interface potential of the substrate which prefers infinitely thick wetting layer.  The Laplace pressure  (or the effective chemical potential $\mu_{\rm eff}$ in Eq.~(\ref{Eq:29x})) that is necessary to produce semi-circular liquid-vapor interface of the droplet becomes positive by the interface potential {even under the undersaturated vapor (Fig.~\ref{fig:5})}.  Then, the bulk coexistence does not play any role to the critical droplet in the complete-wetting regime.  Instead, the prewetting line plays the role of the coexistence in the complete-wetting regime.  Various properties of the critical nucleus changes continuously as functions of the chemical potential.  Therefore the nucleation rate (Eq.~(\ref{Eq:1x})) is expected to changes continuously at the bulk coexistence.  Our result support the conclusion reached from a numerical simulations in Ising system by Sear~\cite{Sear08}. He simply stated that this small nucleus does not "know" whether it will grow to form a wetting layer of finite thickness or a bulk phase of divergent thickness.  Our re-examination of various properties of the critical nucleus in the complete-wetting regime supports his conclusion and, hopefully, will rekindle the interest on heterogeneous nucleation on a completely-wettable substrate.

\begin{acknowledgments}
This work was supported by the Grant-in-Aid for Scientific Research [Contract No.(C)22540422] from Japan Society for the Promotion of Science (JSPS). 
\end{acknowledgments}

\appendix
\section{Continuity of the work of formation $W$ and its derivatives $d^{n}W/d\mu^{n}$at the coexistence $\mu=0$}
Since, the full potential $\phi\left(l,\mu \right)$ is non-singular at $\mu=0$ as shown in Fig.~\ref{fig:4}, all quantities including the solution $l$ of the Euler-Lagrange equations (\ref{Eq:13x}) and (\ref{Eq:47x}) is expected to be non-singular at $\mu=0$.  Therefore, the work of formation $W$ calculated from Eqs.~(\ref{Eq:5x}) and (\ref{Eq:40x}) is a continuous function of the chemical potential $\mu$ at the coexistence $\mu=0$ as shown in Figs.~\ref{fig:7}(a) and \ref{fig:9}.

For a $d=2$ dimensional cylindrical droplet, we can check not only the continuity of $W$ but also that of its derivative $d^{n}W/d\mu^{n}$ directly from Eq.~(\ref{Eq:40x}).  First, the work of formation $W$ in Eq.~(\ref{Eq:40x}) is a continuous function of the chemical potential $\mu$ at $\mu=0$ as indicated by Fig.~\ref{fig:7}(a) since the thin-film thickness $l_{e}\left(\mu\right)$ determined from $d\phi/dl=0$, the droplet height $L\left(\mu\right)$ determined from Eq.~(\ref{Eq:18x}), and the integrand of Eq.~(\ref{Eq:40x})
\begin{equation}
w\left(l, \mu\right)
=\left(2\gamma\Delta\phi\left(l, \mu\right)-\Delta\phi\left(l, \mu\right)^{2}\right)^{1/2}
\label{Eq:A1}
\end{equation}
as a function of the chemical potential $\mu$ are all non-singular at $\mu=0$ (see Eq.~(\ref{Eq:6x}) and Fig.~\ref{fig:4}).

Also, the derivatives $dW/d\mu$  can be continuous (differentiable) at $\mu=0$ as long as the derivatives $dl_{e}/d\mu$, $dL/d\mu$, and $\partial w\left(l, \mu\right)/\partial \mu$ exist and are all non-singular at $\mu=0$.  For example, from $d\phi\left(l_{e}\right)/dl=0$, we find
\begin{equation}
\frac{dV\left(l_{e}\right)}{dl}=\mu.
\label{Eq:A2}
\end{equation}
By differentiating Eq.~(\ref{Eq:A2}), we obtain
\begin{equation}
\left.\frac{dl_{e}}{d\mu}\right|_{\mu=0}=\left(\frac{d^{2}V\left(l_{0}\right)}{dl^{2}}\right)^{-1},
\label{Eq:A3}
\end{equation}
where $l_{0}=l_{e}\left(\mu=0\right)$.  Therefore $dl_{e}/d\mu$ exist at $\mu=0$ and is non-singular as far as the right-hand side of Eq.~(\ref{Eq:A3}) is finite.  By successively differentiating Eq.~(\ref{Eq:A2}), we can easily prove that the $n$-the derivative of thin-film thickness $d^{n}l_{e}/d\mu^{n}$ exists and is given generally by
\begin{equation}
\left.\frac{d^{n}l_{e}}{d\mu^{n}}\right|_{\mu=0}=f_{n}\left( \frac{d^{2}V\left(l_{0}\right)}{dl^{2}} , \dots, \frac{d^{n+1}V\left(l_{0}\right)}{dl^{n+1}}\right)
\label{Eq:A4}
\end{equation}
for $n\ge1$, where $f_{n}$ is a rational function.  Then, $n$-the derivative $d^{n}l_{e}/d\mu^{n}$ will be non-singular at $\mu=0$ as far as all derivatives $d^{k}V\left(l_{0}\right)/dl^{k}$ ($k=1,2,\dots,n+1$) are finite.

A similar argument can be applied to the droplet height $L\left(\mu\right)$ determined from Eq.~(\ref{Eq:18x}).  Again by differentiating Eq.~(\ref{Eq:18x}) and using Eq.~(\ref{Eq:A3}) and $dV\left(l_{0}\right)/dl=0$ from Eq.~(\ref{Eq:A2}), we find
\begin{equation}
\left.\frac{dL}{d\mu}\right|_{\mu=0}=\left(L_{0}-l_{0}\right)\left(\frac{dV\left(L_{0}\right)}{dl}\right)^{-1}
\label{Eq:A5}
\end{equation}
exists and is non-singular at $\mu=0$, where $L_{0}=L\left(\mu=0\right)$.  Again, we can easily prove by mathematical induction that the $n$-th derivative of droplet height $d^{n}L/d\mu^{n}$ is given generally by
\begin{eqnarray}
\left.\frac{d^{n}L}{d\mu^{n}}\right|_{\mu=0}
&=& g_{n}\left( l_{0}, \frac{d^{2}V\left(l_{0}\right)}{dl^{2}} , \dots, \frac{d^{n}V\left(l_{0}\right)}{dl^{n}},\right. \nonumber \\
&& \left.
L_{0}, \frac{dV\left(L_{0}\right)}{dl} , \dots, \frac{d^{n}V\left(L_{0}\right)}{dl^{n}}
\right),
\label{Eq:A6}
\end{eqnarray}
for $n\ge 2$, where $g_{n}$ is a rational function.  Therefore, the derivatives $d^{n}L/d\mu^{n}$ exists and is non-singular at $\mu=0$ if all the arguments of $g_{n}$ in Eq.~(\ref{Eq:A6}) are non-singular. 

Since all the partial derivatives $\partial w^{k}\left(l, \mu\right)/\partial \mu^{k}$  ($k=1,\dots,n$)  are non-singular at $\mu=0$, the derivative $d^{n}W/d\mu^{n}$ will be continuous at $\mu=0$ as far as all the derivatives $d^{k}l_{e}/d\mu^{k}$ and $d^{k}L/d\mu^{k}$ ($k=1,\dots,n$), and, therefore, those of the effective potential $d^{k}V\left(l_{0}\right)/dl^{k}$ ($k=2,\dots,n+1$) at $l_{0}$ and  $d^{k}V\left(L_{0}\right)/dl^{k}$ ($k=1,\dots,n$) at $L_{0}$ are non-singular.   In particular, since our model potential Eq.~(\ref{Eq:19x}) is infinitely differentiable at $l_{0}$ and $L_{0}$, all derivatives $d^{n}l_{e}/d\mu^{n}$ and $d^{n}L/d\mu^{n}$ are expected to be continuous at $\mu=0$ (Fig.~\ref{fig:4}).  Therefore, not only the work of formation $W$ but also its all the derivatives $d^{n}W/d\mu^{n}$ will be continuous at $\mu=0$.

\end{document}